\def\la{\mathrel{\mathchoice {\vcenter{\offinterlineskip\halign{\hfil
$\displaystyle##$\hfil\cr<\cr\sim\cr}}}
{\vcenter{\offinterlineskip\halign{\hfil$\textstyle##$\hfil\cr
<\cr\sim\cr}}}
{\vcenter{\offinterlineskip\halign{\hfil$\scriptstyle##$\hfil\cr
<\cr\sim\cr}}}
{\vcenter{\offinterlineskip\halign{\hfil$\scriptscriptstyle##$\hfil\cr
<\cr\sim\cr}}}}}
 
\def\ga{\mathrel{\mathchoice {\vcenter{\offinterlineskip\halign{\hfil
$\displaystyle##$\hfil\cr>\cr\sim\cr}}}
{\vcenter{\offinterlineskip\halign{\hfil$\textstyle##$\hfil\cr
>\cr\sim\cr}}}
{\vcenter{\offinterlineskip\halign{\hfil$\scriptstyle##$\hfil\cr
>\cr\sim\cr}}}
{\vcenter{\offinterlineskip\halign{\hfil$\scriptscriptstyle##$\hfil\cr
>\cr\sim\cr}}}}}  

\def\p {$\pm$}

\def\kms {\hbox{${\rm km\, s}^{-1}$}} 
\def\cmsq  {$\hbox{{\rm cm}}^{-2}$}    
\def\arcsec {\hbox{$^{\prime\prime}$}}

\def\percc {$\hbox{{\rm cm}}^{-3}$}    
\def\MOLH {\hbox{${\rm H}_2$}}  
\def\MOLN {\hbox{${\rm N}_2$}}  
\def\AMM {\hbox{${\rm NH}_{3}$}} 
\def\HCOP {\hbox{${\rm HCO}^+$}}      
\def\DCOP {\hbox{${\rm DCO}^+$}}    
\def\HTHP {\hbox{${\rm H}_{3}^{+}$}}   
\def\HTDP {\hbox{${\rm H}_{2}{\rm D}^{+}$}}   
\def\HTHOP {\hbox{${\rm H}_{3}{\rm O}^{+}$}} 
\def\NTHP {\hbox{${\rm N}_2{\rm H}^+$}} 
\def\NTDP {\hbox{${\rm N}_2{\rm D}^+$}} 

\documentclass{aa}
\usepackage{graphicx}
\usepackage{amsmath}
\usepackage{dcolumn}

\begin{document}

   \title{Abundant ~\HTDP \ in the pre--stellar core L1544}

   \author{P. Caselli\inst1 \and F. F. S. van der Tak\inst2 \and C. Ceccarelli\inst3
          \and A. Bacmann\inst4 
          }


   \institute{Osservatorio Astrofisico di Arcetri, Largo E. Fermi 5,
        I-50125 Firenze, Italy (caselli@arcetri.astro.it)
         \and 
          Max--Planck--Institut f\"ur Radioastronomie, Auf dem H\"ugel
   	69, 53121 Bonn, Germany (vdtak@mpifr--bonn.mpg.de)
	\and
	Laboratoire d' Astrophysique de l' Observatoire de Grenoble; BP 53,
	38041 Grenoble Cedex, France (cecilia.ceccarelli@obs.ujf--grenoble.fr)
	\and
	European Southern Observatory, Karl--Schwarzschild--Str. 2,
	D--85748 Garching bei M\"unchen, Germany (abacmann@eso.org)   }

   \date{Received on Wed, 19 Feb 2003; accepted }

   \titlerunning{~\HTDP \ in L1544}

\abstract { We have detected the 372 GHz line of {\it ortho}--\HTDP \ 
towards the pre--stellar core L1544. The strongest emission ($T_{\rm mb}$ 
$\sim$ 1 K) occurs at the peak of the millimeter continuum emission, while 
measurements at offset positions indicate that \HTDP \ is confined within 
$\sim$ 20\arcsec , where CO is highly depleted.  The
derived \HTDP \ abundance of $\sim$10$^{-9}$ is comparable with previous 
estimates of the electron abundance in the core, which suggests that \HTDP \ 
is the main molecular ion in the central 20\arcsec \ (2800 AU) of L1544. 
This confirms the expectations that \HTDP \ is dramatically enhanced in gas 
depleted of molecules other than \MOLH . The measured abundance 
even exceeds the present model predictions by about a factor ten.
One possibility is that all CNO--bearing neutral 
species, including atomic oxygen, are almost completely ($\ga$ 98\%) 
frozen within 
a radius of $\sim$2800 AU.
}
\maketitle

\section{Introduction}

The last few years have seen a boom of the studies of
molecular deuteration in star forming regions, triggered
by the discovery of a large fraction of doubly deuterated
formaldehyde in the low mass protostar IRAS16293-2422
(\cite{CCL98}; \cite{LCC00}), where
the measured D$_2$CO/H$_2$CO ratio is 25 times larger than 
in Orion (\cite{T90}).
It is now clear that IRAS16293-2422 is not unusual,
for all Class 0 protostars show similar or larger
D$_2$CO/H$_2$CO ratios (\cite{LCC02}).
Formaldehyde is not unusual either,
as other molecules present astonishing deuteration degrees.
Notable examples are methanol, whose deuterated forms are 
as abundant as the main isotopomer (\cite{PCT02}), or 
ammonia, where not only the doubly deuterated form
(\cite{RTC00}; \cite{LCC01}),
but also the triply deuterated form (\cite{TSM02}; \cite{LRG02}) 
has been detected.

The picture emerging from those studies is that this extreme
molecular deuteration starts during the pre-collapse phase,
when the gas is cold and dense (\cite{CLC01}). 
Whether formed in the gas phase (e.g. \cite{RM00})
or on the grain surfaces (\cite{T83}; \cite{CSS02}),
deuterated isotopomers of neutral species are stored into the 
grain mantles, and released back into the gas phase during the collapse phase,
as the dust is heated by the new born star. This picture has 
been substantially confirmed by the recent observations
of \cite{BLC03}, who measured D$_2$CO/H$_2$CO$\sim 10$\%
in a sample of pre-stellar cores presenting large
degrees of CO depletion (\cite{BLC02}).
A decisive factor for the large observed deuteration seems in fact to be 
the CO depletion: the larger the CO depletion, the larger the 
molecular deuteration.
This suggests that the enhanced molecular deuteration is the product
of gas phase chemistry, given that CO depletion leads to an increase
of the \HTDP /\HTHP \ abundance ratio (\cite{DL84}) and 
thus stimulates the formation of deuterated molecules, mainly produced
in ion--neutral reactions with \HTDP . 

Evidently, the key molecule for fully understanding the molecular
deuteration process is indeed H$_2$D$^+$. 
Unfortunately, the only H$_2$D$^+$ transition observable from ground
based telescopes, the ortho-H$_2$D$^+$ $1_{10}-1_{11}$ transition, 
lies at 372 GHz, next to an atmospheric band
which makes the observation rather difficult.
As a result,
the search of H$_2$D$^+$ has been a very frustrating business for 
about two decades (e.g. \cite{PBK85}; \cite{PWF92}; 
\cite{DPK92}; \cite{BB93}), 
and although several massive protostars have been targetted, 
H$_2$D$^+$ has been so far detected, with a relatively low signal 
(main beam temperature, T$_{\rm mb} \leq 0.1$ K), 
in only two low mass protostars:
NGC1333--IRAS4 (\cite{STD99}) and IRAS16293-2422 (Stark et al., in
preparation).
We decided to re-start the  H$_2$D$^+$ search in sources
with the largest expected H$_2$D$^+$/H$_3^+$ ratio, even where
the excitation conditions of the 372 GHz line may not be the most favorable.
And, since molecular deuteration is indeed dramatically enhanced 
during the pre-collapse phase of low mass protostars, we decided to 
search for H$_2$D$^+$ in the best studied pre-stellar core.

L1544 consists of a dense core surrounded by a low density envelope
which is undergoing extended infall (\cite{TMM98}; \cite{WMW99}).  
From dust emission and absorption observations, we 
know that the core has a central density of about 10$^6$ \percc \ 
inside a radius $r_{\rm flat}$ = 2500 AU (the ``flattened'' region), 
followed by $1/r^2$ density 
fall--off until a radius of about 10000 AU (\cite{WMA99}; 
\cite{BAP00}). Molecular species such as CO, CS, and CCS
are highly depleted at densities above $\sim$ 10$^5$ \percc , or 
inside radii of $\sim$6000 AU (\cite{CWT99}; \cite{OLW99}; 
\cite{TMC02} 2002).  On the other hand, molecules such as
\AMM , \NTHP \ and \NTDP \ do not show signs of depletion, 
probably because of the volatility of their precursor \MOLN \ 
(\cite{BL97}; \cite{CWZ02b}, 
hereafter CWZ).  Indeed, \NTHP \ and \NTDP \ have been 
used to study gas kinematics (\cite{CWZ02c}) and the ionization
degree (CWZ) of the high density core
nucleus. The $N(\NTDP)/N(\NTHP)$  column density ratio 
toward the dust peak was determined to be $\sim$0.2, in agreement with chemical
models which take into account differential depletion of molecular
species in centrally concentrated clouds (\cite{AOI01} 2001; 
CWZ).  These models predict fractional abundances of 
\HTDP \ to be $\la$ 10$^{-10}$, a value which can be tested by the 
present observations. 

\section{Observations and Results}
\label{sobs}
During two nights in October 2002, we observed the dense core L1544 in 
the $1_{10}-1_{11}$ transition of \HTDP \ ($\nu=372.42134$\,GHz), using the 
Caltech Submillimeter Observatory on Mauna Kea. The spectra were taken in
wobbler switching mode, with a chop throw of 300$\arcsec$. The central
position (the peak of the 1.3~mm continuum dust emission map of 
\cite{WMA99}), as well as the 4 positions offset by $\pm 20\arcsec$ in Right
Ascension and Declination were observed. The {\it rms} reached were of the
order of 110\,mK for the central position and 60-90\,mK for the others.
We used as backend an acousto-optical spectrometer with 50\,MHz
bandwidth. The velocity resolution, as measured from a frequency comb
scan, is 0.101\,km\,s$^{-1}$.
The beam efficiency at $\nu=372$\,GHz was measured on Saturn during
the observations, and found
to be $\approx$0.65 for a $22''$ FWHM beam size.
Pointing was monitored every couple of hours and found to 
be better than 3$\arcsec$. 

\begin{figure}[ht!]
 \centering
\resizebox{8cm}{!}{\includegraphics[angle=0,width=8.0cm]{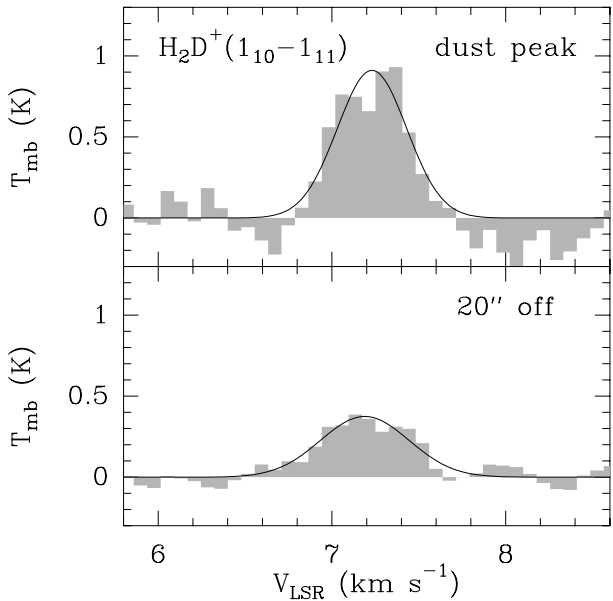}}
 \caption{({\it Top}) The \HTDP (1$_{10}$--$1_{11}$) line at the dust peak of 
L1544 (RA(1950)=05:01:13.1, Dec(1950)=25:06:35.0).  The black curve is the 
Gaussian fit (see Table \ref{tgauss}). 
({\it Bottom}) The \HTDP (1$_{10}$--$1_{11}$) spectrum
averaged in the four positions 20\arcsec \ off the dust peak. }
\label{fspectrum}  
\end{figure}

\begin{figure}[ht!]
 \centering
\resizebox{6.5cm}{!}{\includegraphics[angle=0,width=6.5cm]{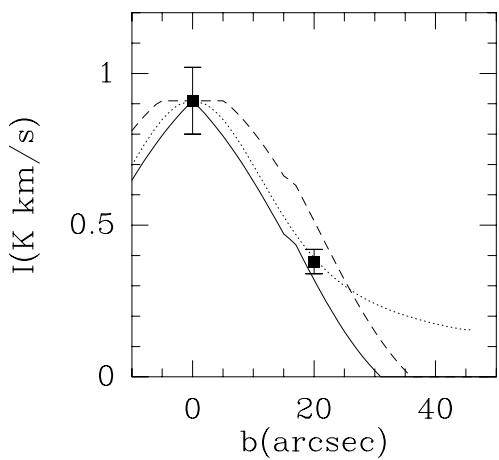}}
 \caption{Integrated intensity of the \HTDP (1$_{10}$--$1_{11}$)
line as a function of projected distance $b$ from the core center.  
The solid and dashed curves show two ``step'' 
functions (null \HTDP \ integrated intensity at impact parameters 
larger than 15 ({\it solid}) and 20 ({\it dashed}) arcsec) convolved with 
the CSO beam. The dotted curve is a (convolved) power--law distribution
in the \HTDP \ integrated intensity ($\propto$ $r^{-1}$), and its size 
at half--maximum (FWHM) is 34\arcsec . }
\label{fconv}  
\end{figure}

In Table \ref{tgauss} we
report the integrated intensity, $V_{\rm LSR}$ velocity, and line 
width of the five observed \HTDP \ spectra.
In Fig.~\ref{fspectrum} the peak spectrum is shown together with the 
average spectrum obtained by summing the four off-peak spectra.  
\begin{table}[tfit]
 \caption{Results of Gaussian fits to the five \HTDP \ spectra}\label{tfit}
 \centering
 \begin{tabular}{lrrr}
 \hline \noalign{\smallskip}
 \multicolumn{1}{c}{Offset} & \multicolumn{1}{c}{$\int T_{\rm mb} dv$} &
\multicolumn{1}{c}{$V_{\rm LSR}$} &
\multicolumn{1}{c}{$\Delta v$} \cr 
\multicolumn{1}{c}{(\arcsec , \arcsec )} & \multicolumn{1}{c}{(K \kms)} & 
\multicolumn{1}{c}{(\kms)} & \multicolumn{1}{c}{(\kms)} \cr
\noalign{\smallskip} \hline \noalign{\smallskip}
(0,0)    & 0.46\p0.04 & 7.23\p0.02 & 0.47\p0.04 \cr
(0,20)   & 0.22\p0.03 & 7.17\p0.04 & 0.49\p0.08 \cr
(-20,0)  & 0.26\p0.04 & 7.11\p0.06 & 0.69\p0.15 \cr
(20,0)   & 0.25\p0.03 & 7.22\p0.03 & 0.59\p0.08 \cr
(0,-20)  & 0.26\p0.04 & 7.26\p0.05 & 0.68\p0.11 \cr
\noalign{\smallskip} \hline \label{tgauss}
\end{tabular}
\end{table}
First, the most striking feature of the \HTDP (1$_{10}$--$1_{11}$) line at the
dust peak is its intensity ($\sim$1 K), which is at least ten times stronger 
than any other previously detected \HTDP \ line (\cite{STD99}),
suggesting that pre--stellar core conditions favour the  production of \HTDP .  The 
second important result is the detection of a strong
signal ($\sim$0.4 K) about 20\arcsec \ from the center.
Assuming that the \HTDP \ emission is constant within a certain radius
$r(\HTDP )$ and then drops to zero, one can estimate $r(\HTDP )$
by convolving the predicted \HTDP \ integrated intensity profile 
with the CSO beam and comparing it with the data.  This comparison,
shown in Fig.~\ref{fconv} (thin and dashed curves), gives $r(\HTDP )$ $\sim$ 
15\arcsec -- 20 \arcsec , consistent with $r_{\rm flat}$ = 17\arcsec , 
the size of the ``flattened'' region observed in the millimeter continuum
emission. In the case of a power law distribution 
of the \HTDP \ integrated intensity ($\int T_{\rm mb} dv  \propto r^{-1}$; 
dotted curve in 
Fig.~\ref{fconv}), the estimated (convolved) size is $r(\HTDP )$ = 17\arcsec . 
This result suggests that the bulk of the \HTDP \ emission is concentrated
{\it within} the highly depleted high density core nucleus. 

The central line profile shows a dip at the rest velocity which 
may be due to the kinematics of the emitting gas (in analogy 
with other molecular ion profiles; see \cite{CWZ02c}) or it may 
indicate that the line is optically thick.  We will get back on this
issue in a future paper where the velocity structure of the L1544 
core will be included in a detailed radiative transfer code (van der 
Tak et al., in preparation).

\section{Column density and abundance}
\label{scolumn}

As shown in Fig.~2, the \HTDP \ emission appears to be 
concentrated within a radius of $\sim$2800 AU.  In this region,
the gas temperature is predicted to be $\sim$ 7 K (\cite{ERS01}; 
\cite{GWG02} 2002), although molecular data suggest larger temperatures
($\sim$ 9--10 K; \cite{TMC02} 2002, \cite{BLC02}). However, because of the 
large amount of gas phase depletion (see 
Sect.~\ref{sdiscussion}), it is likely that molecular tracers do 
not provide information on the gas temperature in the central 2800 AU.  
Assuming LTE conditions ($T_{\rm ex}$ = 7 K), we can 
immediately determine the optical depth ($\tau$) of the 
\HTDP (1$_{10}$--$1_{11}$) line:
\begin{eqnarray}
\tau & = & -ln \left[ 1 - \frac{T_{\rm mb}}{J_{\nu}(T_{\rm ex}) 
	-  J_{\nu}(T_{\rm cb})} \right] \,\, =  \,\, 1.0\pm0.2 ,  
\label{etau}
\end{eqnarray}
where $T_{\rm mb}$ =0.91\p0.11 K at the peak, $T_{\rm cb}$ = 2.7 K
is the cosmic background temperature, and $J_{\nu}(T)$ = 
$T_0/(exp(T_0/T)-1)$ with $T_0$ = $h\nu/k$ = 17.9 K.
Assuming a spontaneous transition coefficient 
$A$ = 1.08$\times$10$^{-4}$ s$^{-1}$,  the total 
column density of ortho--\HTDP \ is:
\begin{eqnarray}
N_{\rm ortho} & \simeq & 5.15\,\, 10^{13} \, \Delta v(\kms ) \,\,
\tau \,\, {\rm cm^{-2}}
\end{eqnarray} 
so that at the dust peak, where $\tau \sim$1 and the linewidth 
is 0.47 \kms (Table \ref{tgauss}), we 
obtain $N_{\rm ortho}$ = 2.4$\times$10$^{13}$ \cmsq .  At the low assumed
temperatures, where the ortho to para ratio is close 
to 1 (\cite{GHR02} 2002), we deduce a 
total \HTDP \ column density of $\sim$4.8$\times$10$^{13}$ \cmsq , 
about three times larger than $N(\NTHP )$ (see CWZ).

To gauge the \HTDP \ abundance, the above estimate of the
total \HTDP \ column density has been divided by the molecular 
hydrogen column density derived from the 1.3~mm continuum emission 
of Ward--Thompson et al. (1999), smoothed at a resolution of 
22\arcsec \ ($N(\MOLH )$=1.1$\times$10$^{23}$ \cmsq ): 
$x(\HTDP )$ $\equiv$ $N(\HTDP )/N(\MOLH) \sim 4.4\times 10^{-10}$.
This value is only four 
times lower than the estimated electron abundance $x(e)$ in 
the L1544 center ($\sim$ 2$\times$10$^{-9}$, CWZ), suggesting
that \HTDP \ is one of the major ions in the gas phase (see 
Sect.~\ref{sdiscussion}). 

In case of subthermal emission, 
the line optical depth and the total column density increase.  
For example, if $T_{\rm ex}$ = 6 K, $\tau$ = 2.8 and 
the total \HTDP \ column density becomes $\sim$ 1.2$\times$10$^{14}$ 
\cmsq .  This corresponds to $x(\HTDP )$ $\sim$ 1$\times$10$^{-9}$, 
so that $x(\HTDP )$ $\simeq$ $x(e)$. We note that $T_{\rm ex}$ = 6 K is 
the minimum excitation temperature possible to have a solution of 
eq.~(\ref{etau}) in the case of an optically thick ($\tau \ga$ 0.5) 
line with $T_{\rm mb}$ = 0.9 K.  
At the off--positions (20\arcsec \ or 2800 AU from the dust peak), 
where $N(\MOLH )$ $\sim$ 8$\times$10$^{22}$ \cmsq , the derived 
\HTDP \ abundance is 2.3$\times$10$^{-10}$ (if $T_{\rm ex}$ 
= 7 K) and 4.0$\times$10$^{-10}$ (if $T_{\rm ex}$ = 6 K), 
a factor of about two lower than toward the dust peak.  

We also carried out a radiative transfer analysis using the Monte Carlo 
program by \cite{HT00}\footnote{http://talisker.as.arizona.edu/$\sim$michiel/ratran.html}. 
The \MOLH \ density profile was taken from \cite{TMC02} (2002), 
whereas dust temperatures are from \cite{GWG02} (2002). Only thermal line 
broadening was included. Term
energies, statistical weights and Einstein A coefficients were taken
from the JPL catalog (\cite{pick98}).
Besides collisional excitation, for which we used the scaled radiative
rates proposed by \cite{BDW90}, radiation from the cosmic
microwave background and thermal radiation by local dust are taken
into account, using grain opacities from \cite{OH94} (1994).
The abundance of \HTDP \ was assumed to be constant within a radius $r_0$
and zero at $r > r_0$. Model results are compared to line profiles at the
center and offset positions to constrain the \HTDP \ abundance and $r_0$.
The best-fit abundance is \HTDP /\MOLH = 10$^{-9}$, where calibration uncertainty 
allows a 50\% decrease.  Models with $r_0$ = 20--25 arcsec give good matches to 
the line fluxes at both the center and the offset positions. 
The optical depth at line
center along the central pencil beam is 2.73. Here we limited our analysis 
to a static cloud.
In a future paper (van der Tak et al., in prep.) we will 
explore the effects of including a velocity structure in the radiative
transfer code.

\section{Chemistry}
\label{sdiscussion}

The main result of this study is that \HTDP \ has been detected
for the first time in a pre--stellar core.  The observed  
($1_{10}-1_{11}$) line has a main beam temperature of $\sim$ 1 K,
and the emission is  concentrated in a region with 
radius of $\sim$2800 AU, roughly coincident with the so--called 
``flattened region'' seen in the millimeter dust continuum 
emission, where the density is slowly changing with radius
($\sim r^{-1.4}$; see \cite{AWB00}) and/or the temperature
is around 7 K (\cite{ERS01}; \cite{GWG02} 2002).  

The data have been analysed with a simple analytical model and 
with a Monte Carlo radiative transfer code.  The derived \HTDP \ abundance
($x(\HTDP )$ $\sim$ 10$^{-9}$) is consistent 
with the electron fraction estimated by CWZ in the center of 
L1544. This implies that \HTDP \ is the main molecular ion 
in the core center.  The simple chemical code of CWZ, 
constrained by observations of several molecular ions, predicts 
$x(\HTDP )$ $\sim$ 7$\times$10$^{-11}$ in the center of L1544, 
about one order of magnitude less than that deduced from the present
data.  A similar \HTDP \ abundance ($\sim$ 6$\times$10$^{-11}$) at the
cloud center was also predicted by the detailed chemical model 
of \cite{AOI01} (2001), coupled with the Larson--Penston dynamical
evolution which best reproduced the L1544 
observational results.

CWZ found that to match the 
observed \NTDP /\NTHP \ ($\sim$0.24) and \DCOP /\HCOP \ ($\sim$0.06) 
column density ratios 
toward the L1544 center one needs to (i) allow differential depletion
of molecular species onto dust grains (in particular, \MOLN \ has to
be more volatile than CO, to maintain a large fraction of \NTHP \ 
and \NTDP \ in the core center); and 
(ii) maintain a significant fraction of atomic oxygen in the gas 
phase, to limit the deuterium fractionation to the observed values
($\la$ 0.2).
As a by--product, \HTHOP \ was predicted to be abundant ($x(\HTHOP )$
$\sim$ 10$^{-9}$) in the core center.   

The results of the present paper apparently contradict the 
CWZ conclusions, because the large \HTDP \
abundance can only be produced if all neutral species, including
O and \MOLN, are essentially frozen onto dust grains (see also 
\cite{BAH02}). In fact, only in this case can the \HTDP \ abundance 
reach values close to the electron abundance.
This can be seen with a simple chemical model,  where the
ingredients are H, \MOLH , HD, \HTDP , 
\HTHP , CO, \HCOP , \DCOP and electrons. 
We neglect recombination onto 
dust grains, which is likely not to be an important process for the 
above molecular ions in 
the high density core nucleus, where small grains and PAHs -- significant
carriers of negative charges (\cite{LD88}) --   are expected 
to be deposited onto bigger grains (e.g. \cite{OH94} 1994).  
Following section 2 of \cite{C02} (2002), assuming conditions appropriate 
for the central 2500 AU of L1544 (kinetic temperature = 7 K, $n({\MOLH })$ = 
10$^{6}$ \percc ,  see \cite{GWG02} 2002), we have:
\begin{eqnarray}
R_{\rm DEUT} \equiv \frac{x(\HTDP )}{x(\HTHP )} & = & 
\frac{2\, 10^{10} \, k_{\rm f}}{3\, 10^2 / f_{\rm D} \, +  \, 
 x(e)/2\, 10^{-9}}
\label{erd}
\end{eqnarray}
where $k_{\rm f}$ is the forward rate coefficient of the proton--deuteron 
exchange reaction \HTHP  +HD $\rightleftharpoons$ \HTDP +\MOLH \ (see below),
and $f_{\rm D}$ is defined as the ratio between 
the ``canonical'' CO abundance (9.5$\times$10$^{-5}$;
\cite{FLW82}) and the observed $N({\rm CO})/N(\MOLH )$.  
The \HTHP \ abundance can be estimated from the charge conservation
equation, so that we obtain (with rate coefficients as in Caselli 2002):
\begin{eqnarray}
x(\HTHP ) & = & \frac{f_{\rm D} x(e)^2}{(f_{\rm D} x(e) + 4 \, 10^{-8})
(R_{\rm DEUT} + 1) + 8 \, 10^{-8}}.
\label{eh3p}
\end{eqnarray}
Equations (\ref{erd}) and (\ref{eh3p}) 
directly furnish the \HTDP \ and \HTHP \ abundances as a function of 
$f_{\rm D}$, once $x(e)$ is known. 
Figure 3 shows the result of this simple chemical analysis, with 
$x(\HTDP )$ (thin curves) and $R_{\rm DEUT}$ (dashed curves) as a function
of $f_{\rm D}$, assuming $x(e)$ = 2$\times$10$^{-9}$ (as roughly determined
by CWZ in the L1544 center).  We note that eqn.~(\ref{erd}) breaks down 
at large values of $f_{\rm D}$ ($\ga$ 100), when dissociative
recombination starts to dominate over proton transfer reactions.  
In the figure, two different values for $k_{\rm f}$  
have been used: (i) the ``standard'' value $k_{\rm f}$ = 
1.5$\times$10$^{-9}$ cm$^3$ s$^{-1}$ (upper curves), and 
(ii) $k_{\rm f}$ = 3.5$\times$10$^{-10}$ cm$^3$ s$^{-1}$,
recently  measured by \cite{GHR02} (2002) (lower curves).
The two data points in the figure are our present estimates of $x(\HTDP )$,
the arithmetic means of the abundances calculated with 
$T_{\rm ex}$=6 and 7 K (Sect.~\ref{scolumn}): 7.2$\times$10$^{-10}$ and 
3.15$\times$10$^{-10}$ at the peak and off--peak positions, respectively.  
The corresponding $f_{\rm D}$ values are from CWZ. 
The error bars indicate the uncertainties
in the excitation temperature ($\sim$1 K; Sect.~\ref{scolumn}) for 
estimating $x(\HTDP )$, and calibration errors affecting the determination 
of $f_{\rm D}$ (about 30\%; CWZ).  

\begin{figure}[ht!]
 \centering
 \resizebox{8cm}{!}{\includegraphics[angle=0,width=8.0cm]{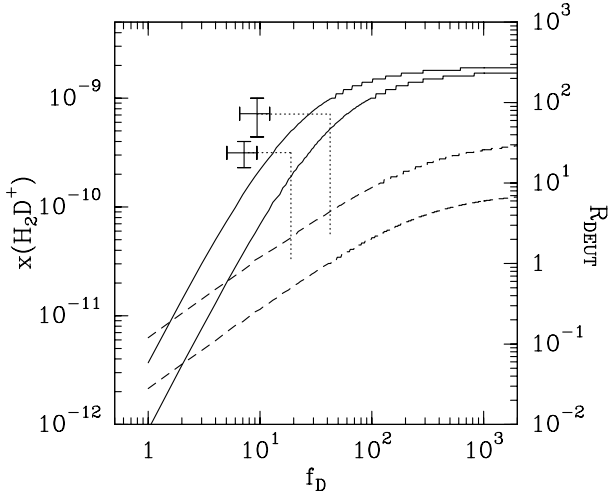}}
 \caption{({\it thin curves}) $x(\HTDP )$ vs. the depletion 
factor $f_{\rm D}$ in the high density core nucleus of 
L1544 (see text). ({\it dashed curves}) $R_{\rm DEUT}$ as a 
function of $f_{\rm D}$. Data points are the measured values 
at the peak and 20\arcsec \ off--peak positions. Dotted lines indicate
the $R_{\rm DEUT}$ and $f_{\rm D}$ values predicted by the simple 
chemical model applied to the L1544 central 2800 AU 
(see Section\ref{sdiscussion}), at the observed $x(\HTDP )$ 
values. Large \HTDP \ abundances ($\geq 10^{-9}$) are obtained only if 
$f_{\rm D} \geq 70$.}
\label{fchemistry}  
\end{figure}

The observed \HTDP \ abundances cannot be reproduced unless the 
CO depletion factor is 
$\sim$20 in the off position and $\sim$40 at the peak (the ``Monte Carlo
abundance'' indicates an even higher $f_{\rm D}$, $\simeq$70), whereas the 
observed $f_{\rm D}$ values are $\sim$7 and $\sim$9, at the same positions.  
Although our comprehension of the deuteration process in cold 
gas may be substantially incomplete, it is certain that measurements 
of $f_{\rm D}$ are contaminated by the CO emission along the 
line of sight coming from regions with marginal CO depletion, so that 
the resultant $f_{\rm D}$'s estimates are
lower limits of the total amount of CO freeze out at the core center.  
Indeed, the chemical model of CWZ predicts an 
almost complete ($\ga$99\%) CO freeze out within the central 2800 AU of L1544.
Following the model results in Fig.~\ref{fchemistry}, the predicted 
$f_{\rm D}$ values at the observed \HTDP \ abundances correspond to 
$R_{\rm DEUT}$ $\sim$ 1 and 2 at the off and central positions, respectively.

If other neutrals reacting 
with \HTDP \ and \HTHP , such as O and \MOLN , are present in the 
gas phase, the predicted $x(\HTDP )$ abundance will drop as in the 
case of a low $f_{\rm D}$ value (e.g. \cite{C02} 2002).  
Thus, the present \HTDP \ observations
also require a heavy depletion of O ($\ga$99\%) {\it and} 
\MOLN \ ($\sim$97\%).  A way to overcome the apparent inconsistency with 
CWZ (who need significant fractions of  molecular nitrogen and  
atomic oxygen in the 
gas phase at $r$ $\sim$ 2800 AU) and well reproduce both the moderate value 
($\sim$ 0.2) of the $N(\NTDP )/N(\NTHP )$ column density ratio 
as well as the large \HTDP \ abundance (which implies $R_{\rm DEUT}$ $\sim$
2) is to increase the depletion rate of O and \MOLN \ {\it within} the 
``flattened'' region.  This case requires central core densities 
of about 10$^{7}$ \percc \ within a radius of 1000 AU, which is consistent 
with current mm continuum dust emission measurements if the central 
temperature drops to values of $\la$ 7 K (e.g.
\cite{ERS01}).  The conclusion is that present observations 
are compatible with {\it \HTDP \ being the most abundant molecular ion
and with a total depletion of elements heavier than helium in the central 
$\la$ 2800 AU}.  This can also explain the 
lower \HTDP \ abundance derived in the direction of NGC1333--IRAS4
($x(\HTDP )$ = 3$\times$10$^{-12}$; \cite{STD99}), where 
the young stellar object heats up the central zones, thus allowing 
the return of solid phase molecules back into the gas phase.
A more comprehensive chemical model,
coupled with a detailed radiative transfer code is however
needed to better understand the whole data set available for 
L1544 and we are going to explore this in a future paper (van 
der Tak et al., in preparation). 
\begin{acknowledgements}
The authors are grateful to Malcolm Walmsley for discussion and to Dominic 
Benford for his assistance during the observations.  The CSO is supported by 
NSF grant AST 99-80846.
PC acknowledges support from the MIUR project ``Dust and Molecules in 
Astrophysical Environments''.  
\end{acknowledgements}

\end{document}